\documentclass[pdftex,twocolumn,epjc3,preprintnumbers,amsmath,amssymb]{svjour3}          % twocolumn

\RequirePackage[T1]{fontenc}

\smartqed  % flush right qed marks, e.g. at end of proof

\RequirePackage{graphicx}
\RequirePackage{mathptmx}   % use Times fonts if available on your TeX system
\RequirePackage{flushend}
\RequirePackage[numbers,sort&compress]{natbib}
\RequirePackage[colorlinks,citecolor=blue,urlcolor=blue,linkcolor=blue]{hyperref}

\journalname{Eur. Phys. J. C}
\usepackage{amsmath}
\usepackage{amssymb}
\usepackage[normalem]{ulem}
\usepackage{bm}
\usepackage{color}
\usepackage{graphicx}
\usepackage[colorlinks,citecolor=blue,urlcolor=blue,linkcolor=blue]{hyperref}
\usepackage{multirow}
\usepackage{subfigure}
\usepackage{booktabs}
\usepackage{mathrsfs}
\usepackage{balance}
\newcommand{\f}{\frac}
\newcommand{\lt}{\left}
\newcommand{\m}{m_{\rm P}}
\newcommand{\n}{\nonumber}
\newcommand{\p}{\partial}
\newcommand{\rt}{\right}
\newcommand{\dd}{{\rm d}}
\newcommand{\bt}{\beta}
\newcommand{\dt}{\delta}

\newcommand{\ve}{\varepsilon}
\newcommand{\sg}{\sigma}

\newcommand{\pb}{{\rm PBH}}

\begin{document}
%\onecolumn
\title{Primordial black holes from the ultraslow-roll phase \\ in the inflaton--curvaton mixed field inflation} %\thanksref{t1}}


\author{Bing-Yu Su\thanksref{addr1,addr2} 
\and 
Nan Li\thanksref{addr3,e1} 
\and 
Lei Feng\thanksref{addr1,addr2,e2}} 

\thankstext{e1}{e-mail: linan@mail.neu.edu.cn (corresponding author)}
\thankstext{e2}{e-mail: fenglei@pmo.ac.cn (corresponding author)}

\institute{Key Laboratory of Dark Matter and Space Astronomy, Purple Mountain Observatory, Chinese Academy of Sciences, Nanjing 210023, China\label{addr1} 
\and 
School of Astronomy and Space Science, University of Science and Technology of China, Hefei, Anhui 230026, China\label{addr2} 
\and 
Department of Physics, College of Sciences, Northeastern University, Shenyang 110819, China\label{addr3} }

\date{Received: date / Accepted: date}

\maketitle

\begin{abstract}
Primordial black holes (PBHs) are a promising candidate for dark matter, as they can form in the very early universe without invoking new particle physics. This work explores PBH formation within a curvaton scenario featuring an ultraslow-roll (USR) phase. An inflaton--curvaton mixed field model is presented, where the inflaton drives early inflation and then transits into the USR phase, amplifying the small-scale curvature perturbation. During inflation, the curvaton generates entropy perturbation, which later converts into curvature perturbation after the curvaton decays in the radiation-dominated era. Using the $\delta N$ formalism, we compute the power spectrum of the total primordial curvature perturbation and analyze the relevant non-Gaussianity. Our results show that adding a curvaton field not only has a significant impact on primordial non-Gaussianity, but also introduces more complex inflationary dynamics, even saving the inflaton potentials that generate too low scalar spectral indices. Our model can produce PBHs with mass around $10^{-14}\,M_\odot$ that account for all dark matter, while remaining consistent with current observational constraints.

\end{abstract}

\section{Introduction} \label{sec:intro}

Primordial black holes (PBHs) are hypothesized to form in the very early radiation-dominated era of the universe. Among the various mechanisms proposed so far \cite{Carr:2020gox, Green:2020jor, Carr:2021bzv}, the simplest one involves the gravitational collapse of the regions where density fluctuations exceed a certain threshold, leading to the direct formation of PBHs from the radiation field \cite{Zeldovich:1967lct, Hawking:1971ei, Carr:1974nx, Carr:1975qj}. Unlike the black holes formed from stellar processes, PBHs have a mass range that spans from the Planck mass to supermassive scales, covering dozens of orders of magnitude. Due to the Hawking radiation, the PBHs with masses $M_{\rm PBH}<10^{-18}\,M_\odot$ ($M_\odot=2\times 10^{33}$ g is the solar mass) have already evaporated, while those with larger masses can still exist today. 

PBHs have become increasingly significant in cosmology due to their potential roles in a variety of phenomena. They can influence the cosmic ray background through the Hawking radiation \cite{Boudaud:2018hqb, Laha:2019ssq, Arbey:2019vqx, laha2020n, Calabrese:2021zfq, Huang:2024xap, Su:2024hrp, DelaTorreLuque:2024qms}, help explain some gravitational wave events \cite{LIGOScientific:2016sjg, LIGOScientific:2017bnn, Sasaki:2018dmp}, and, if supermassive, contribute to cosmic structure formation through the Poisson or seed effects \cite{Hoyle1966, 1975A&A....38....5M, 1983ApJ...268....1C, Carr1984, Carr:2018rid}. More importantly, PBHs are also considered as a promising candidate for dark matter (DM) \cite{Green:2020jor, Carr:2020xqk, Carr:2021bzv}. The PBH abundance $f_\pb$ refers to the fraction of the current DM density that is attributed to PBHs. If $f_\pb \gtrsim 0.1$, PBHs can effectively account for a substantial portion of DM. Currently, various observations have placed stringent constraints on $f_\pb$ across different mass ranges \cite{Carr:2020gox, Escriva:2022duf}, leaving one possible mass window at $10^{-16}$--$10^{-11}\,M_\odot$, where PBHs may still constitute the entirety of DM. 

To produce a sufficient abundance of PBHs, the power spectrum $\mathcal{P}_\zeta$ of the primordial curvature perturbation must undergo a substantial amplification, increasing from $10^{-9}$ on the cosmic microwave background (CMB) scale \cite{Planck:2018vyg, Planck:2018jri} up to $10^{-3}$--$10^{-2}$ on much smaller scales. Such a significant enhancement can be realized in different theoretical frameworks, including single- \cite{Kannike:2017bxn, Ballesteros:2017fsr, Gangopadhyay:2021kmf} and multi-field inflation models \cite{Gordon:2000hv, Cai:2021wzd, Hooshangi:2022lao, Wilkins:2023asp, Qin:2023lgo}. In single-field inflation models, the typical slow-roll (SR) conditions are insufficient to generate such large $\mathcal{P}_\zeta$ on small scales. However, this can be realized through an ultraslow-roll (USR) phase \cite{Garcia-Bellido:2017mdw, Motohashi:2017kbs, Dimopoulos:2017ged, Di:2017ndc, Ozsoy:2018flq, Cheng:2018qof, Bhaumik:2019tvl, Mishra:2019pzq, Ozsoy:2020kat, Liu:2020oqe, Ragavendra:2020sop, Figueroa:2020jkf, Zhang:2021vak, Wang:2021kbh, Liu:2021qky, Cheng:2021lif, Figueroa:2021zah, Wu:2021mwy, Mu:2022dku, Ragavendra:2023ret, Zhao:2023zbg, Zhao:2023xnh, Zhao:2024yzg, Biswas:2023jcd}. During the USR phase, the inflaton field $\phi$ experiences a period of extremely slow motion due to the presence of a bump, dip, or plateau on the background inflaton potential, leading to a huge increase in $\mathcal{P}_\zeta$ and consequently resulting in a substantial abundance of PBHs. The USR condition can be achieved in many ways, such as imposing one or two perturbations on $V(\phi)$ \cite{Ozsoy:2018flq, Mishra:2019pzq, Ozsoy:2020kat, Zhang:2021vak, Wang:2021kbh, Liu:2021qky, Zhao:2023zbg, Zhao:2023xnh, Zhao:2024yzg} or constructing an inflaton potential $V(\phi)$ with a (near-)inflection point \cite{Garcia-Bellido:2017mdw, Ragavendra:2020sop, Figueroa:2020jkf, Cheng:2021lif, Figueroa:2021zah}. The USR mechanism can also be adapted to multi-field inflation models \cite{Hooshangi:2022lao}, also leading to fruitful discussions \cite{Fu:2022ssq, Tomberg:2023kli}.

A well-known multi-field approach for PBH formation is the curvaton scenario \cite{Linde:1996gt, Enqvist:2001zp, Lyth:2001nq, Moroi:2001ct, Bartolo:2002vf, Lyth:2002my, Bartolo:2003jx, Sasaki:2006kq, Huang:2008ze, Beltran:2008aa, Pi:2021dft, Lodman:2023yrc}, with the inflaton--curvaton mixed field model \cite{Legrand:1995xc, Langlois:2004nn, Langlois:2008vk, Fonseca:2012cj, Byrnes:2014xua, Chen:2019zza, Meng:2022low, Chen:2023lou, Yu:2023jrs} being a prominent example. In such model, two scalar fields, the inflaton $\phi$ and the curvaton $\chi$, operate independently. On the one hand, the inflaton $\phi$ governs the cosmic expansion during inflation and creates the curvature perturbation $\zeta_\phi$. On the other hand, the curvaton $\chi$, which is subdominant during inflation and decoupled from the inflaton, generates the primordial entropy (isocurvature) perturbation ${S}_\chi$. To comply with the stringent CMB constraints, ${S}_\chi$ must remain smaller than the dominant curvature perturbation $\zeta_\phi$ \cite{Enqvist:2000hp, Enqvist:2001zp}. Once the universe enters the radiation-dominated era, the curvaton behaves like pressure-less matter. As the Hubble expansion rate decreases and eventually drops below the curvaton mass $m_\chi$, the curvaton begins to oscillate. Before the onset of Big Bang nucleosynthesis (BBN), the curvaton has completely decayed into radiation, converting its entropy perturbation into curvature perturbation entirely \cite{Mollerach:1989hu}. As a consequence, this process provides an alternative mechanism for enhancing $\mathcal{P}_\zeta$ and potentially triggering PBH formation. 

Our work synthesizes the two proposals discussed above and constructs an inflaton--curvaton mixed field model with a USR phase to produce the PBHs with a typical mass $M_{\rm PBH}$ around $10^{-14}\,M_\odot$. In this model, the inflaton drives inflationary dynamics. After the early SR phase, it transits into a USR phase so as to enhance the small-scale curvature perturbation. Simultaneously, the curvaton generates entropy perturbation during inflation. After its decay, the entropy perturbation ultimately transform into curvature perturbation. With the curvature perturbation produced by the inflaton, these perturbations together influence the structure formation across various scales significantly. Furthermore, the nonlinear nature of the conversion between the entropy and curvature perturbations inevitably introduces non-Gaussianity into the total curvature perturbation $\zeta$ \cite{Sasaki:2006kq, Ferrante:2022mui}. %This non-Gaussianity, particularly from the curvaton with very low decay fraction, can produce a heavy-tailed distribution of $\zeta$. 
In this work, we employ the $\delta N$ formalism to analyze the perturbations introduced by both the inflaton and curvaton fields, and then demonstrate how these perturbations evolve into the final curvature perturbation $\zeta$, which influences the PBH abundance, by investigating two types of inflaton potentials featuring the USR phases within our model.

The structure of this paper is as follows. In Sec. \ref{sec:model}, we develop the inflaton--curvaton mixed field model with a USR phase and calculate the resulting curvature perturbation. In Sec. \ref{sec:nongau}, we analyze the CMB constraints on the primordial non-Gaussianity. Section \ref{sec:mf} covers the calculation procedure of the PBH mass and abundance. Finally, in Sec. \ref{sec:contri}, we explore the power spectrum and PBH abundance with the PBH mass around $10^{-14}\,M_\odot$ in two different kinds of mixed field inflation models with the USR phase. We conclude and discuss our results in Sec. \ref{sec:con}. In this paper, we work in the natural system of units and set $c=\hslash=k_{\rm B}=1$.

\section{Inflaton--curvaton fixed field Model} \label{sec:model}
%{The power spectrum of the mixed field model}

In this section, we construct an inflaton--curvaton mixed field inflation model with a USR phase, and then utilize the $\dt N$ formalism to discuss the power spectrum of the total primordial curvature perturbation in detail.

\subsection{Inflationary dynamics}

We adopt the action for the inflaton--curvaton mixed field inflation model as
\begin{align}
S(\phi,\chi)&=\int\dd^4 x\,\sqrt{-g}\n\\
&\quad\times\lt[\f{m_{\rm P}^2}{2}R-\f{1}{2}(\p\phi)^2-\frac{1}{2}(\p\chi)^2-V_{\rm tot}(\phi,\chi)\rt], \n
\end{align}
where $m_{\rm P}=1/\sqrt{8\pi G}$ is the reduced Planck mass, $R$ is the Ricci scalar, and $V_{\rm tot}(\phi,\chi)$ is the total potential of the inflaton $\phi$ and curvaton $\chi$ fields, respectively. Both the two fields $\phi$ and $\chi$ can be decomposed into the backgrounds and perturbations as $\phi=\bar\phi+\dt\phi$ and $\chi=\bar\chi+\dt\chi$. In this work, we assume that the inflaton and curvaton fields evolve independently, so $V_{\rm tot}(\phi,\chi)$ can be expressed as the sum of their individual potential. We further consider a simple quadratic potential for the curvaton field, so
\begin{align}
V_{\rm tot}(\phi,\chi)=V(\phi)+\f{1}{2}m_\chi^2\chi^2,\label{Vtot}
\end{align}
where $V(\phi)$ is the inflaton potential with a USR phase, and $m_\chi$ is the curvaton mass. 

To describe the motion of the inflaton in the cosmic expansion, two SR parameters are introduced as
\begin{align}
\ve&=-\frac{\dot H}{H^2}=\frac{{\bar\phi}_{,N}^2}{2\m^2}, \n\\
\eta&=-\frac{\ddot{\bar\phi}}{H\dot{\bar\phi}}=\frac{{\bar\phi}_{,N}^2}{2m_{\rm P}^2}-\frac{{\bar\phi}_{,NN}}{{\bar\phi}_{,N}}, \n
\end{align}
%where the symbol $\dot{}$ denotes differentiation with respect to physical time $t$, while the subscripts ${,x}$ represents the derivative with respect to $x$. 
where $H=\dot a/a$ is the Hubble expansion rate, $a$ is the scale factor, and $N$ is the number of $e$-folds, defined as $\dd N=H\,\dd t=\dd\ln a$. During the SR inflation, $\ve,|\eta|\ll 1$, but in the USR phase, these two conditions remarkably break down. In addition, in order to study the acceleration of the inflaton $\phi$ and curvaton $\chi$ separately, we introduce two new parameters defined by the total potential $V_{\rm tot}(\bar\phi,\bar\chi)$ of the two scalar fields as 
\begin{align}
\eta_\phi&=\f{m_{\rm P}^2V_{,\bar\phi\bar\phi}(\bar\phi)}{V_{\rm tot}(\bar\phi,\bar\chi)},\n\\
\eta_\chi&=\f{m_{\rm P}^2m_\chi^2}{V_{\rm tot}(\bar\phi,\bar\chi)}. \n
\end{align}
In the SR phase, we have $|\eta_\chi|\ll|\eta_\phi|\approx|\eta|$. However, in the USR phase, $|\eta_\phi|$ may deviate from $|\eta|$ slightly. Throughout inflation, $|\eta_\chi|$ keeps almost constant. 

On the homogeneous and isotropic background, the equations of motion for the inflaton and curvaton fields are described by the Klein--Gordon equations as
\begin{align}
\ddot{\bar\phi}+3H\dot{\bar\phi}+V_{,\bar\phi}(\bar\phi)&=0,\label{phimo}\\
\ddot{\bar\chi}+3H\dot{\bar\chi}+m_\chi^2\bar\chi&=0.\label{chimo}
\end{align}
It is worth noting that the curvaton mass should be very small, satisfying $m_\chi\ll H$ during inflation. As a result, on the super-Hubble scales, the curvaton is effectively frozen, so $H$ is approximately given by the Friedmann equation as
\begin{align}
H^2&=\f{1}{3m_{\rm P}^2}\lt[\f 12{\dot{\bar\phi}}^2+\f 12{\dot{\bar\chi}}^2+V_{\rm tot}(\bar\phi,\bar\chi)\rt]\n\\
&\approx \f{1}{3m_{\rm P}^2}\lt[\f 12{\dot{\bar\phi}}^2+V(\bar\phi)\rt] \n\\
&=\frac{V(\bar\phi)}{(3-\ve)m^2_{\rm P}}, \n
\end{align}
and Eq. (\ref{phimo}) can be rewritten via the number of $e$-folds $N$ as
%in terms of the SR parameter $\ve$ and the $e$-folds $N$ as
\begin{align}
{\bar\phi}_{,NN}+(3-\ve){\bar\phi}_{,N}+\frac{V_{,\bar\phi}(\bar\phi)}{H^2}&=0. \n
\end{align}
%Additionally, to conveniently describe cosmic expansion, we use the number of $e$-folds $N$ as the time variable. This is defined by $\dd N=H\,\dd t=\dd\ln a$, which allows us to analyze the evolution of the inflaton field $\phi$. Here $a$ is the scale factor and the Hubble expansion rate $H=\dot{a}/a$. A sufficient inflationary phase of at least 60--70 $e$-folds is required to solve the flatness and horizon problems in the standard hot Big Bang model, typically not exceeding 80

Next, we turn to the perturbations on the background universe. During inflation, the inflaton induces the curvature perturbation $\zeta_\phi$, and the curvaton produces the entropy perturbation ${S}_\chi$. %These perturbations are assumed to be Gaussian. After inflation but before the onset of BBN, the curvaton field decays into radiation, and the entropy perturbation ${S}_\chi$ finally converts into the curvature perturbation $\zeta_\chi$. 
Moreover, for the metric perturbation, since the vector and tensor perturbations are irrelevant to PBH formation, we focus on the scalar perturbation $\Phi$, and the perturbed metric in Newtonian gauge reads 
\begin{align}
\dd s^2=-(1+2\Phi)\,\dd t^2+a^2(t) (1-2\Phi)\dt_{ij}\,\dd x^i\dd x^j. \n
\end{align}

In the mixed field inflation model, the total comoving curvature perturbation $\cal R$ is \cite{Gordon:2000hv}
\begin{align}
\mathcal{R}=\Phi+H\f{\dot{\bar\phi}\,\dt\phi+\dot{\bar\chi}\,\dt\chi}{{\dot{\bar\phi}}^2+{\dot{\bar\chi}}^2}. \n %=\Phi-\f{H}{\dot H}(\dot\Phi+H\Phi)
\end{align}
During inflation, we assume that the contribution of the curvaton field to the energy density is negligible, so we have %the comoving curvature perturbation is approximately
\begin{align}
\mathcal{R}\approx\Phi+H\f{\dt\phi}{\dot{\bar\phi}}. \label{R}
\end{align}
The corresponding equation of motion for $\mathcal{R}_{k}$ in the Fourier space is the Mukhanov--Sasaki equation \cite{Sasaki:1986hm, Mukhanov:1988jd},
\begin{align}
{\cal R}_{k,NN}+(3+\ve-2\eta){\cal R}_{k,N}+\frac{k^2}{H^2e^{2N}}{\cal R}_k=0. \n
\end{align}

Now, we move on to the power spectra of the perturbations. First, for the curvature perturbation $\zeta_\phi=-H\,\dt\phi/\dot{\bar\phi}$ caused by the inflaton, from Eq. (\ref{R}), we know that $\zeta_\phi\approx-\mathcal{R}$ on the uniform-energy-density slice in the large-scale limit where $k\to 0$. Thus, during the SR phase, its power spectrum $\mathcal{P}_{\zeta_\phi}$ is nearly scale-invariant and can be approximated using the SR parameters as
\begin{align}
\mathcal{P}_{\zeta_\phi}\approx\f{1}{8\pi^2\ve}\lt(\f{H_*}{m_{\rm P}}\rt)^2, \label{Pzetaphislow}
\end{align}
where $*$ denotes the epoch of the Hubble-exit. However, during the USR phase, because $\zeta_\phi$ will continue to evolve on the super-Hubble scales, we have
\begin{align}
\mathcal{P}_{\zeta_\phi}\approx\mathcal{P}_{\mathcal{R}}=\f{k^3}{2\pi^2}|\mathcal{R}_k|^2\Bigg|_{k\ll aH},\label{Pzetaphi}
\end{align}
which must be evaluated at the end of inflation.

Second, on a spatially-flat slice, the curvaton perturbation $\dt\chi$ satisfies the equation of motion as 
\begin{align}
\dt\ddot{\chi}_k+3H\dt{\dot{\chi}}_k+\lt(\f{k^2}{a^2}+m_\chi^2\rt)\dt\chi_k=0. \label{dtchi}
\end{align}
Therefore, on the super-Hubble scales where $k\to 0$, the equations of motion for $\chi$ and $\dt\chi$ in Eqs. (\ref{chimo}) and (\ref{dtchi}) coincide, so $\dt\chi\propto\bar\chi$ holds until the curvaton decays \cite{Lyth:2001nq, Lyth:2002my, Kohri:2012yw}. Thus, for convenience, we define the curvaton density contrast as 
\begin{align}
\dt=\f{\dt \chi}{\bar\chi}. \n
\end{align} 
%During inflation, the vacuum fluctuations of a light scalar field on the sub-Hubble scales are stretched to the super-Hubble scales by the accelerated expansion, leading to a perturbation power spectrum. 
Like the inflaton, the curvaton perturbation originates mainly from quantum fluctuation as $\dt\chi\approx H/(2\pi)$ \cite{Wands:2002bn, Fonseca:2012cj}. On the super-Hubble scales and on the spatially-flat slice, neglecting the SR corrections and the correlations between the inflaton and curvaton perturbations \cite{Byrnes:2006fr}, the power spectrum of $\dt\chi$ can be approximated as
\begin{align}
\mathcal{P}_{\dt\chi_*}\approx \lt(\f{H_*}{2\pi}\rt)^2. \n
\end{align}
Hence, the power spectrum of $\dt$ at the end of inflation is%\footnote{Attention, $\mathcal{P}_{\dt}$ is not to be confused with $\mathcal{P}_{\dt\chi}$.}
\begin{align}
\mathcal{P}_{\dt}%=\f{k^3}{2\pi^2}|\dt_{\chi k}|^2\Bigg|_{\rm end}
\approx\lt(\f{H_*}{2\pi\bar\chi_*}\rt)^2. \label{Pdtchi}
\end{align}
As can be easily seen, ${\cal P}_\delta$ depends on scales very weakly. 

\subsection{Post-inflation dynamics} \label{sec:spectrum}

On the super-Hubble scales, the curvature perturbation on the uniform-energy-density slice can be obtained via the $\dt N$ formalism as \cite{Bartolo:2003jx, Lyth:2004gb, Sasaki:2006kq}
\begin{align}
\zeta_i(t,{\bf x})=\dt N(t,{\bf x})+\f{1}{3}\int^{\rho_i(t,{\bf x})}_{\bar\rho_i(t)}\f{\dd \tilde\rho_i}{\tilde\rho_i+p_i(\tilde\rho_i)},\label{dtN}
\end{align}
where $\rho_i$ and $\bar\rho_i$ represent the energy density and its background value of the $i$-th component (such as the inflaton field $\phi$, the curvaton field $\chi$, or the total energy density), and $p_i$ denotes the pressure of the $i$-th component. Here, 
\begin{align}
\dt N(t,{\bf x})= N(t,{\bf x})-\bar N(t) \n %=\zeta(t,{\bf x})
\end{align} 
is the perturbation of the number of $e$-folds $N$. The $\dt N$ formalism helps analytically compute the evolution of the large-scale curvature perturbation, without needing to solve the complicated perturbative equation. 
%According to Eq. (\ref{dtN}), the local energy density of the $i$-th component can be written as
%\begin{align}
%\rho_i=\bar\rho_i e^{3(1+w_i)(\zeta_i-\dt N)},\label{rhoi}
%\end{align}
%where $w_i=p_i/\rho_i$ is the equation of state parameter.

In the radiation-dominated era, when the Hubble expansion rate drops to $H\ll m_\chi$, the curvaton begins to oscillate around the minimum of its potential. Before the curvaton decays, it behaves like pressure-less matter, so $\bar\rho_\chi\propto a^{-3}$. From Eq. (\ref{dtN}), we have
%For the oscillating curvaton after inflation, its energy density can be expressed as
\begin{align}
\rho_\chi=\bar\rho_\chi e^{3(\zeta_\chi-\dt N)}, \label{rhochi}
\end{align}
where $\zeta_\chi$ is the conserved curvature perturbation from the curvaton field, and $\dt N=\zeta_\phi$ on the spatially-flat slice \cite{Langlois:2008vk}. Eq. (\ref{rhochi}) can be rewritten as $\rho_\chi=\bar{\rho}_\chi e^{{S}_\chi}$, where 
\begin{align}
{S}_\chi= 3(\zeta_\chi - \zeta_\phi)\n
\end{align}
is the entropy perturbation from the curvaton field. 
%During the post-inflationary phase, the curvaton does not dominate the cosmic background evolution, so it is represented in a spatially-flat slice as $\dt N=\zeta_\phi$ \cite{Langlois:2008vk}. Thus, we find
%\begin{align}
%\rho_\chi=\bar\rho_\chi e^{3(\zeta_\chi-\zeta_\phi)}=\bar\rho_\chi e^{{S}_\chi}.\label{rhochiphi}
%\end{align}
%where
%\begin{align}
%{S}_\chi\equiv3(\zeta_\chi-\zeta_\phi)\label{Schi}
%\end{align}
%represents the entropy perturbation carried by the curvaton perturbation $\dt\chi$. 
Moreover, the time-averaged energy density of the curvaton field is $\rho_\chi=m_\chi^2 \chi^2=m_\chi^2 (\bar\chi+\dt\chi)^2=\bar\rho_\chi(1+\dt)^2$, with $\bar\rho_\chi=m_\chi^2 \bar\chi^2$ \cite{Lyth:2002my, Langlois:2008vk}. This leads to $e^{{S}_\chi}=(1+\dt)^2$, which can be further expanded to the linear order as
\begin{align}
{S}_\chi
%=3(\zeta_{\chi}-\zeta_{\phi})
=2\dt_{\chi}%+\dt_{\chi}^2+\mathcal{O}(\dt_{\chi})^4={S}_{\chi,{\rm G}}+\f{1}{4}{S}_{\chi,{\rm G}}^2
.\label{Schi2}
\end{align}
%where ${S}_{\chi,{\rm G}}$ denotes the linear order of the entropy perturbation, representing the Gaussian component of ${S}_\chi$.

Assuming that, at the end of inflation, all the inflaton energy decays into radiation, the resulting inflaton perturbation will generate the curvature perturbation in the radiation energy density $\rho_{\rm R}$ \cite{Langlois:2008vk}. Before the curvaton decays, from Eq. (\ref{dtN}), the radiation energy density is
\begin{align}
\rho_{\rm R}=\bar\rho_{\rm R}e^{4(\zeta_\phi-\dt N)}.\label{rhoR}
\end{align}
For convenience, we further assume that the curvaton decays instantaneously when $H\sim t_{\rm dec}^{-1}$, where $t_{\rm dec}$ is the lifetime of the curvaton and is assumed as a constant. These assumptions yield the results consistent with the numerical simulations in Ref. \cite{Malik:2006pm}. Afterward, $\dt N$ can be taken as the total curvature perturbation $\zeta$ produced by both the inflaton and curvaton fields. If we suppose that all curvaton decay products are relativistic, $\zeta$ remains conserved after the curvaton decay but before the horizon re-entry. Hence, the background total energy density on a uniform-total-energy-density slice at $t_{\rm dec}$ is
\begin{align}
\bar\rho_{\rm rad}(t_{\rm dec})=
\rho_{\rm R}(t_{\rm dec},{\bf x})+\rho_\chi(t_{\rm dec},{\bf x})=\f{3m_{\rm P}^2}{t_{\rm dec}^2}.\label{rhotot}
\end{align}

From Eqs. (\ref{rhochi}), (\ref{rhoR}), and (\ref{rhotot}), we obtain
\begin{align}
e^{3(\zeta_\chi-\zeta)}\Omega_{\chi}(t_{\rm dec})+e^{4(\zeta_\phi-\zeta)}[1-\Omega_{\chi}(t_{\rm dec})]=1,\label{zetarela}
\end{align}
where $\Omega_{\chi}(t_{\rm dec})=\bar\rho_\chi/(\bar\rho_{\rm R}+\bar\rho_{\chi})$ is the curvaton density fraction at $t_{\rm dec}$. At the linear order, Eq. (\ref{zetarela}) provides the total curvature perturbation as
\begin{align}
\zeta=(1-r)\zeta_{\phi}+r \zeta_{\chi}=\zeta_{\phi}+\f{2r}{3}\dt_{\chi}, \label{zetabbb1}
\end{align}
where $r\in(0,1)$ is the curvaton decay fraction at $t_{\rm dec}$, defined as \cite{Lyth:2001nq}
\begin{align}
r=\f{3\Omega_{\chi}(t_{\rm dec})}{4-\Omega_{\chi}(t_{\rm dec})}=\f{3\bar\rho_\chi}{4\bar\rho_{\rm R}+3\bar\rho_\chi}\Bigg|_{t_{\rm dec}}. \n
\end{align}
%with $0<r<1$. %Assuming that the inflaton-generated curvature perturbation is Gaussian, the Gaussian component of the total curvature perturbation is
%\begin{align}
%\zeta%=\zeta_{\phi}+\f{r}{3}{S}_{\chi}
%=\zeta_{\phi}+\f{2r}{3}\dt_{\chi}.\label{zetab1dt}
%\end{align}

Given that the inflaton-induced curvature perturbation $\zeta_\phi$ and the curvaton-induced entropy perturbation ${S}_\chi$ are uncorrelated, after the curvaton decay, the power spectrum $\mathcal{P}_{\zeta}$ of the total curvature perturbation $\zeta$ can be approximated as \cite{Fonseca:2012cj}
\begin{align}
\mathcal{P}_{\zeta}
\approx%\mathcal{P}_{\zeta_{\phi}}+\f{r^2}{9}\mathcal{P}_{{S}_{\chi}}=
\mathcal{P}_{\zeta_\phi}+\f{4r^2}{9}\mathcal{P}_{\dt}. \label{Pzeta}
\end{align}
Using Eqs. (\ref{Pzetaphi}), (\ref{Pdtchi}), and (\ref{Pzeta}), we are able to obtain the final expression of $\mathcal{P}_{\zeta}$,
\begin{align}
\mathcal{P}_\zeta%&=\f{k^3}{2\pi^2}|\mathcal{R}_k|^2\Bigg|_{k\ll aH}+\f{2k^3r^2}{9\pi^2}|\dt_{\chi k}|^2\Bigg|_{\rm end}\n\\&\approx
=\f{k^3}{2\pi^2}|\mathcal{R}_k|^2\Bigg|_{k\ll aH}+\lt(\f{rH_*}{3\pi\bar\chi_*}\rt)^2.\label{Pzetanumb}
\end{align}

To quantify the contribution of the curvaton field relative to the inflaton field, a dimensionless parameter $\lambda$ can be introduced as \cite{Vernizzi:2005fx, Langlois:2008vk, Chen:2019zza}
\begin{align}
\lambda=\f{4r^2\mathcal{P}_{\dt}}{9\mathcal{P}_{\zeta_\phi}}. 
\end{align}
Consequently, $\mathcal{P}_\zeta$ can be rewritten as $\mathcal{P}_{\zeta}=(1+\lambda)\mathcal{P}_{\zeta_\phi}$. Using the SR approximation in Eq. (\ref{Pzetaphislow}), Eq. (\ref{Pzetanumb}) can be further simplified as
\begin{align}
\mathcal{P}_{\zeta}=\lt[\f{1+\lambda}{8\pi^2\ve_*}\lt(\f{H_*}{m_{\rm P}}\rt)^2\rt]\Bigg|_{k\ll k_{\rm PBH}}. \n
\end{align}
Therefore, on large scales, from Eq. (\ref{Pzetaphislow}), the parameter $\lambda$ should be
\begin{align}
\lambda=\f{8\ve_* r^2m_{\rm P}^2}{9\bar\chi_*^2}.\label{lambdaL}
\end{align}

According to the observational constraints on the large CMB scales, ${\cal P}_\zeta$ is typically expressed in a nearly scale-invariant power-law form as
\begin{align}
{\cal P}_{\zeta}^{\rm CMB}(k)=A_{\rm s}\lt(\f{k}{k_{\rm p}}\rt)^{n_{\rm s}-1}, \label{observe} 
\end{align}
where $A_{\rm s}=2.10\times 10^{-9}$ is the amplitude, $n_{\rm s}$ is the scalar spectral index, and $k_{\rm p}=0.05\,{\rm Mpc}^{-1}$ is the CMB pivot scale \cite{Planck:2018jri}. 
% and the corresponding $N_{\rm p}$ is around 30 at the Hubble-exit.
To constrain the parameter space, we focus on $n_{\rm s}$ and the tensor-to-scalar ratio $r_{\rm ts}$, which can be expressed in terms of the SR parameters as \cite{Wands:2002bn, Fonseca:2012cj, Lodman:2023yrc}
\begin{align}
n_{\rm s}-1&=\f{1}{1+\lambda}(2\eta_\phi-6\ve)+\f{\lambda}{1+\lambda}(2\eta_\chi-2\ve), \label{ns}\\
r_{\rm ts}&= \f{16\ve}{1+\lambda}. \label{rts}
\end{align}
%where
%\begin{align}
%n_\phi-1&=2\eta_\phi-6\ve, \n\\
%n_\chi-1&=2\eta_\chi-2\ve.\n
%\end{align}
On the CMB pivot scale $k_{\rm p}$, $n_{\rm s}=0.9649\pm0.0042$ at 68\% CL \cite{Planck:2018vyg, Planck:2018jri}, and $r_{\rm ts}$ is bounded by $r_{\rm ts}<0.036$ at 95\% CL \cite{BICEP:2021xfz}. From Eqs. (\ref{ns}) and (\ref{rts}), we find that the curvaton influences $n_{\rm s}$ and $r_{\rm ts}$ via $\lambda$ and $\eta_\chi$. Without the curvaton field, Eqs. (\ref{ns}) and (\ref{rts}) reduce to $n_{\rm s}-1=2\eta_\phi-6\ve$ and $r_{\rm ts}=16\ve$ as in the case of single-field inflation. Especially, we highlight that adding the curvaton field can significantly increase the scalar spectral index $n_{\rm s}$, to be discussed in detail in Sec. \ref{sec:power}.

Last, to prevent the curvaton from inducing a second inflationary phase, it must remain subdominant compared with radiation before its decay \cite{Langlois:2004nn, Dimopoulos:2011gb}.
% Additionally, assuming weak coupling between the inflaton and curvaton fields, the quantum fluctuations of the curvaton can be treated as a Gaussian distribution \cite{Sasaki:2006kq}. 
Assuming that the quantum fluctuation of the curvaton field is Gaussian \cite{Sasaki:2006kq}, at the Hubble-exit, the curvaton field $\bar{\chi}_*$ should satisfy the condition \cite{Enqvist:2001zp, Langlois:2004nn}
\begin{align}
H_*\ll \bar\chi_*\ll m_{\rm P}.\label{limit}
\end{align}
%We define parameters $h \equiv H_*/m_{\rm P}$ and $m \equiv m_{\rm P}/\bar{\chi}_*$. 
Based on this condition and the observational constraints on the power spectrum on the CMB scale in Eq. (\ref{observe}), we have $H_*/m_{\rm P}<1.3\times10^{-5}$ and $10^{-5}\ll \bar{\chi}_*/m_{\rm P}\ll1$. 

\section{Non-Gaussianity} \label{sec:nongau}
%{Non-Gaussianity of primordial perturbations} 

To further constrain the parameters in our mixed field inflation model, we consult the non-Gaussianity from the CMB observations. Currently, the {\it Planck} full-mission CMB temperature and $E$-mode polarization maps have placed stringent bounds on the non-Gaussianity of primordial curvature perturbation $\zeta$ on large scales \cite{Planck:2019kim}. If the curvaton decay fraction $r$ is too low, significant non-Gaussianity can arise \cite{Lyth:2002my, Bartolo:2003jx, Langlois:2004nn, Fonseca:2012cj, Sasaki:2006kq, Chen:2019zza, Ferrante:2022mui, Hooper:2023nnl, Yu:2023jrs} and thus deserves deeper investigation.

Usually, non-Gaussianity is characterized by the local nonlinear parameter $f_{\rm NL}$ defined as
\begin{align}
\zeta=\zeta_{\rm G}+\f 35f_{\rm NL}\zeta_{\rm G}^2, \n
\end{align}
where $\zeta_{\rm G}$ is Gaussian component of $\zeta$. The CMB temperature and polarization data provide the following constraint \cite{Planck:2019kim},
\begin{align}
f_{\rm NL}=-0.9\pm5.1 \quad (68\%\,{\rm CL}), \label{nnll} %,\quad g_{\rm NL}=(-5.8\pm6.5)\times10^4\,(68\%\,{\rm CL})
\end{align}
suggesting that the level of non-Gaussianity on large scales is very small, consistent with the predictions from single-field SR inflation models. However, it could be more significant on smaller scales, especially on those relevant to PBHs. On these scales, non-Gaussianity may significantly influence the clumping of matter and the formation of PBHs. Unfortunately, current observational data are still insufficient to provide precise constraints on the non-Gaussianity on such scales. %This limitation arises primarily from the weaker signals of non-Gaussianity on small scales, which are heavily affected by noise, complicating the extraction of these signals from existing CMB data.

Generally speaking, the local nonlinear parameter $f_{\rm NL}$ can be formulated as
\begin{align}
f_{\rm NL}=\lt(\f{5}{4r}-\f{5}{3}-\f{5r}{6}\rt)\alpha({\bf k}_1,{\bf k}_2,{\bf k}_3),\n
\end{align}
where the scale-dependent factor $\alpha({\bf k}_1,{\bf k}_2,{\bf k}_3)$ accounts for the relative sizes of the momenta in different configurations of the triangle formed by the vectors ${\bf k}_1$, ${\bf k}_2$, and ${\bf k}_3$, such as the squeezed configuration ($k_1 \approx k_2 \gg k_3$), the equilateral configuration ($k_1 \approx k_2 \approx k_3$), and the folded configuration ($k_1 \approx 2k_2 \approx 2k_3$) \cite{Komatsu:2009kd}. Here, we select the squeezed one, which can be produced by multi-field inflation. The parameter $f_{\rm NL}$ for the inflaton--curvaton mixed field model with a quadratic curvaton potential is given by \cite{Langlois:2004nn, Fonseca:2012cj, Chen:2019zza, Yu:2023jrs}
\begin{align}
f_{\rm NL}=\lt(\f{5}{4r}-\f{5}{3}-\f{5r}{6}\rt)\lt(\f{\lambda}{1+\lambda}\rt)^2.\label{fnl}%\\
%g_{\rm NL}&=\f{25}{54}\lt(-\f{9}{r}+\f{1}{2}+10r+3r^2\rt)\lt(\f{\lambda}{1+\lambda}\rt)^3.\label{gnl}
\end{align}
From Eqs. (\ref{lambdaL}) and (\ref{fnl}), on the large scales where $k_1, k_2, k_3\ll k_{\rm PBH}$, we have
\begin{align}
f_{\rm NL}=\lt(\f{5}{4r}-\f{5}{3}-\f{5r}{6}\rt)\lt(\f{8\ve_* r^2m_{\rm P}^2}{9\bar\chi_*^2+8\ve_* r^2m_{\rm P}^2}\rt)^2.\label{fNLL}%\\
%g_{\rm NL}&=\f{25}{54}\lt(-\f{9}{r}+\f{1}{2}+10r+3r^2\rt)\lt(\f{8\ve_* r^2m^2}{9+8\ve_* r^2m^2}\rt)^3\Bigg|_{k_{\rm p}}.\label{gNLL}
\end{align}
Note that, on the PBH scales, the situation is much more complex, and no reliable approximation method exists for studying the non-Gaussianity in this regime. Therefore, we focus solely on the case with $k_1, k_2, k_3\ll k_{\rm PBH}$ and assume that $r$ is independent of $\bar\chi_*$. 

\begin{figure}[htb]
\centering \includegraphics[width=1\linewidth]{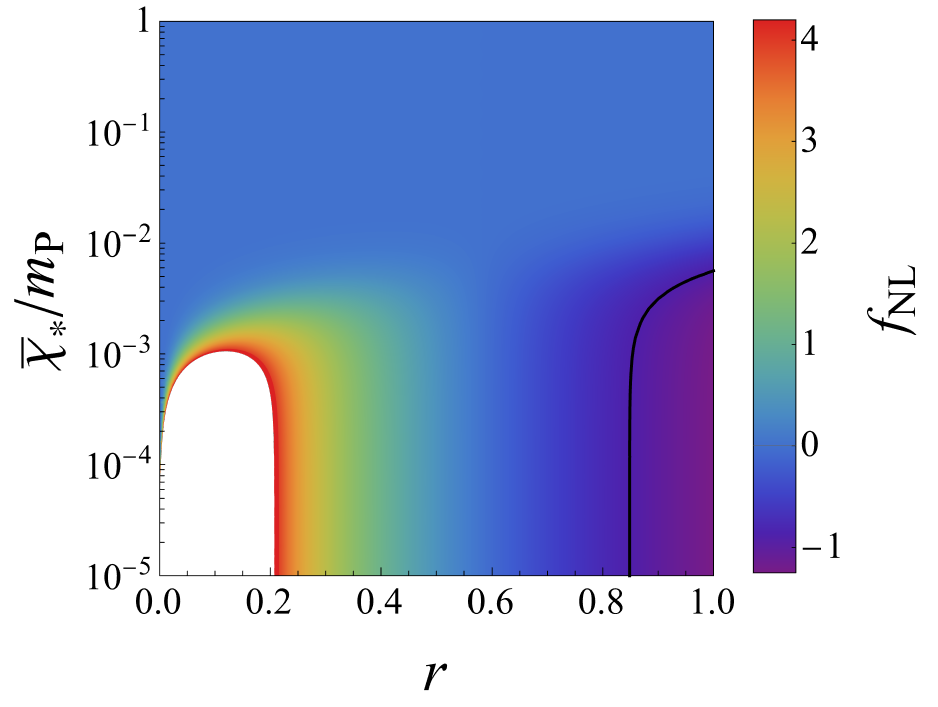}
\caption{The influence of different values of $r$ and $\bar\chi_*$ on the local nonlinear parameter $f_{\rm NL}$. We take $\ve_*\sim2\times10^{-4}$ for illustration, demonstrating that the non-Gaussianity becomes significant at very low values of $r$ and $\bar\chi_*$ (the white region with $r\lesssim 0.2$ and $\bar\chi_*/m_{\rm P}\lesssim 10^{-3}$). This observation further constrains the model parameters in our work. The solid black line corresponds to the case when $f_{\rm NL}=-0.9$.} 
\label{fig:rm}
\end{figure}

In Fig. \ref{fig:rm}, we show the parameter space for $r$ and $\bar\chi_*$ according to Eq. (\ref{fNLL}), where different colors correspond to different values of $f_{\rm NL}$, and the black line is for $f_{\rm NL} = -0.9$. Our results indicate that, when the values of $r$ and $\bar\chi_*$ are too low ($r\lesssim 0.2$ and $\bar\chi_*/m_{\rm P}\lesssim 10^{-3}$), the corresponding level of non-Gaussianity increases and may even exceed the observational limits in Eq. (\ref{nnll}). Therefore, care should be taken in choosing parameters to avoid this issue. Here, we take $\ve_*\sim2\times 10^{-4}$ at the Hubble-exit for the large scales.
%according to the calculation result.

\section{PBH mass and abundance}\label{sec:mf}

When the primordial curvature perturbation with large amplitudes on certain scales re-enters the Hubble horizon during the radiation-dominated era, the resulting overdense regions can collapse to PBHs. The PBH mass $M_{\rm PBH}$ is related to the comoving wavenumber of the curvature perturbation. According to the Carr--Hawking collapse model \cite{Carr:1974nx}, $M_{\rm PBH}$ is given by \cite{Carr:2020gox, Mishra:2019pzq}
\begin{align}
\f{M_{\rm PBH}}{M_{\odot}}=1.13\times10^{15}\lt(\f{\gamma}{0.2}\rt)\lt(\f{g}{106.75}\rt)^{-1/6}\lt(\f{k_{\pb}}{k_{\rm p}}\rt)^{-2}, \n %\approx\lt(\f{k_{\pb}}{1.68\times10^6\,{\rm Mpc}^{-1}}\rt)^{-2}
\end{align}
where $\gamma$ is the collapse efficiency, $k_{\pb} = aH$ is the comoving wavenumber of the PBHs at the Hubble-exit, and $g$ is the effective relativistic degree of freedom for energy density at PBH formation. Below, we take $\gamma=0.2$ and $g=106.75$ \cite{Carr:1975qj}. 

The PBH mass fraction $\beta_{\rm PBH}$ at the time of its formation is defined as
\begin{align}
\beta_{\rm PBH}=\f{\rho_{\rm PBH}}{\rho_{\rm rad}} \Bigg|_{\rm formation}. \n
\end{align}
Based on the Press--Schechter formalism, we obtain \cite{Ando:2018qdb}
\begin{align}
\beta_{\rm PBH} %=\gamma\int_{\dt_{\rm c}}\f{\dd \dt}{\sqrt{2\pi\sg_k^2(R)}}\exp{\lt(-\f{\dt^2}{2\sg_k^2(R)}\rt)}
\approx\f{\gamma\sg_k}{\sqrt{2\pi}\dt_{\rm c}}\exp{\lt(-\f{\dt_{\rm c}^2}{2\sg_k^2}\rt)},\n
\end{align}
where $\dt_{\rm c}$ is the threshold of the radiation density contrast required for PBH formation, and we take $\dt_{\rm c}=0.45$ \cite{Musco:2004ak}. In the momentum space, $\sg_k^2$ represents the variance of the smoothed density over a smoothing scale $R$,
\begin{align}
\sg_k^2=\f{16}{81}\int_0^\infty\f{\dd q}{q}\,\widetilde W^2(q,R)T^2(q,R)(qR)^4\mathcal{P}_\zeta(q), \n
\end{align}
where $\widetilde W(k,R)$ is the Fourier transform of the real-space window function $W(r,R)$ \cite{Ando:2018qdb, Young:2019osy, Tokeshi:2020tjq}. If we choose a simple real-space top-hat window function as $W(r,R)=(4\pi R^3/3)^{-1} \Theta(R-r)$, with $\Theta(x)$ being the Heaviside step function, we have
\begin{align}
\widetilde W(k,R)=3\f{\sin(kR)-(kR)\cos(kR)}{(kR)^3}.\n
\end{align}
The normalization of $W(r,R)$ is $\int\,\dd^3r\,W(r,R)=1$, ensuring $\widetilde W(k=0,R)=1$. The transfer function $T(k,R)$ describes the evolution of the sub-Hubble modes during the radiation-dominated era, which is expressed as
\begin{align}
T(k,R)=3\f{\sin(kR/\sqrt{3})-(kR/\sqrt{3})\cos(kR/\sqrt{3})}{(kR/\sqrt{3})^3}.\n
\end{align}
Note that, when using either Gaussian or the momentum-space top-hat window function, the transfer function becomes irrelevant, as the density perturbation coarse-grained at the horizon scale is insensitive to the sub-Hubble modes. In the study of PBH formation, the smoothing scale $R$ is usually chosen as the comoving Hubble radius $(aH)^{-1}$. 

Finally, for the large-mass PBHs that have not evaporated yet, their present abundance $f_\pb$ is defined as
\begin{align}
f_\pb=\f{\Omega_\pb}{\Omega_{\rm DM}}\Bigg|_{\rm today},\n
\end{align}
where $\Omega_\pb$ and $\Omega_{\rm DM}$ are the energy density parameters of PBHs and DM, respectively. 
% The PBH abundance $f_\pb$ provides a direct link between the early universe perturbation and the present DM content.
Assuming a monochromatic mass distribution for PBHs, $f_\pb$ is naturally proportional to the mass fraction $\beta_\pb$ as \cite{Sasaki:2018dmp, Carr:2020gox}
\begin{align}
f_\pb&=1.68\times 10^{8} %\lt(\f{\gamma}{0.2}\rt)^{1/2}\lt(\f{g}{106.75}\rt)^{-1/4}\n\\&\quad\times
\lt(\f{M_{\rm PBH}}{M_\odot}\rt)^{-1/2} \bt_\pb. \n
\end{align}

\section{PBHs from the mixed field inflation models} \label{sec:contri}

With all above preparations, we can now explore the formation of PBHs within a targeted mass range. In this section, we introduce two different kinds of mixed field inflation models with the USR phase and discuss their relevant physical influences on PBHs. 

\subsection{Two mixed field inflation models} \label{sec:power}

In this subsection, we introduce two different mixed field inflation models. To realize the USR phase, the first one involves a perturbation on the background inflaton potential, and the second one possesses (near-)inflection points.

\subsubsection{Case 1} \label{sec:case1}

First, following our previous works in Refs. \cite{Wang:2021kbh, Liu:2021qky, Zhao:2023zbg, Zhao:2023xnh, Zhao:2024yzg},
We concentrate on the inflaton potential $V_{1}(\phi)$ as
\begin{align}
V_{1}(\phi)=V_{\rm b}(\phi)+\dt V(\phi),\label{V1}
\end{align}
where $V_{\rm b}(\phi)$ and $\dt V(\phi)$ represent the background inflaton potential and its perturbation. We choose $V_{\rm b}(\phi)$ as the Kachru--Kallosh--Linde--Trivedi potential \cite{Kachru:2003aw},
\begin{align}
V_{\rm b}(\phi)=V_0\f{{\phi}^2}{{\phi}^2+(m_{\rm P}/2)^2}, \label{Vb} 
\end{align}
where $V_0$ describes the energy scale of inflation, and then construct an anti-symmetric perturbation $\dt V(\phi)$ as 
\begin{align}
\dt V(\phi)/V_0=-A(\phi-\phi_0)\exp\lt[-\frac{(\phi-\phi_0)^2}{2\sg^2}\rt],\label{dtV}
\end{align}
where $A$, $\phi_0$, and $\sg$ are the parameters characterizing the slope, position, and width of $\dt V(\phi)$, respectively. By substituting Eqs. (\ref{V1})--(\ref{dtV}) into Eq. (\ref{Vtot}), we can obtain the total potential $V_{\rm tot}(\phi,\chi)$. For simplicity, we set $A=V_{{\rm b},\phi}(\phi_0)$ in order to create a perfect plateau around $\phi_0$. Moreover, the initial conditions for inflation are chosen as ${\phi}_{\rm i}/m_{\rm P}=3.30$ and ${\phi}_{{\rm i},N}/m_{\rm P}=-0.0137$.

Below, we adopt the parameters in our mixed field inflation model as $\phi_0/m_{\rm P}=1.625$, $\sg/m_{\rm P}=0.06372$, $V_0/m_{\rm P}^4=8\times10^{-11}$, $r=0.4$, and $\bar\chi_*/m_{\rm P}=10^{-2}$, respectively. These values are selected to guarantee the consistency with both the large-scale observational constraints on $A_{\rm s}$, $n_{\rm s}$, $r_{\rm ts}$, and $f_{\rm NL}$ and the required enhancement of the PBH abundance $f_\pb$ with the desire PBH mass $M_\pb$ on small scales.

\begin{figure}[htb]
\centering 
\includegraphics[width=1\linewidth]{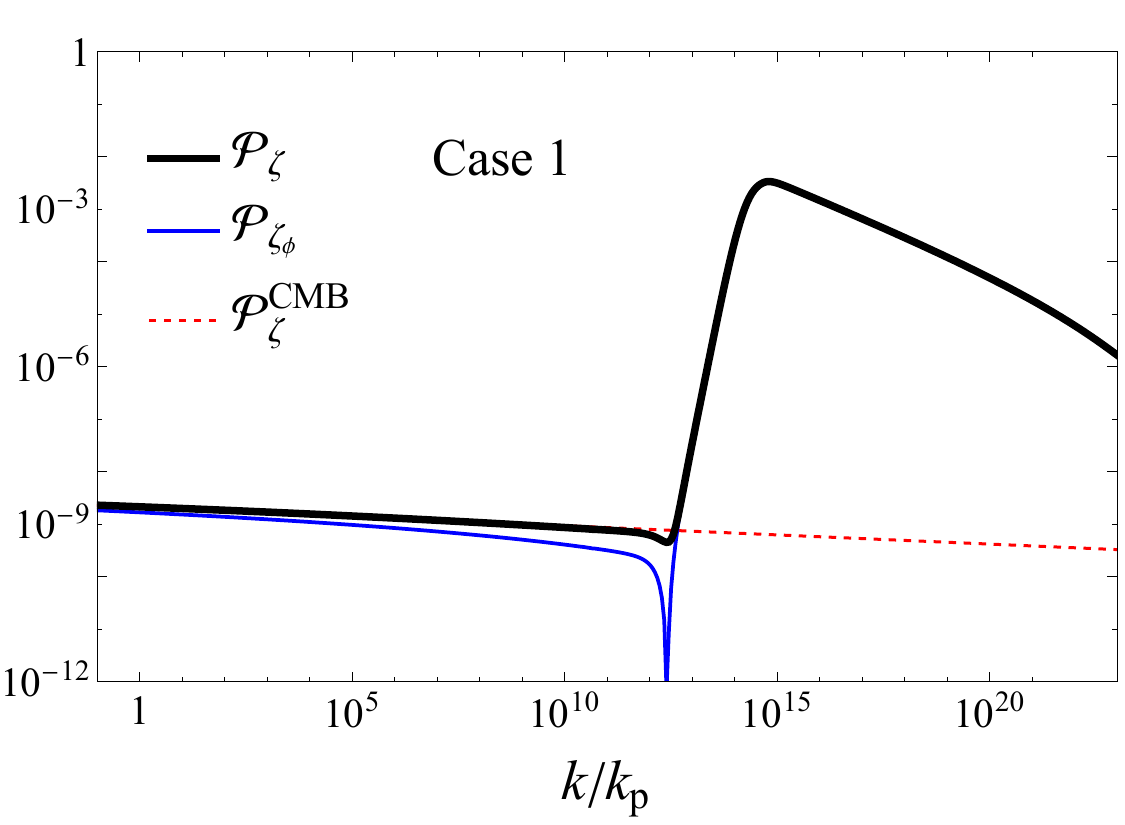}
\caption{The power spectra of total curvature perturbation $\mathcal{P}_\zeta$ (the black line), of only inflaton-induced curvature perturbation $\mathcal{P}_{\zeta_\phi}$ (the blue line), and extrapolated from the CMB observations on large scales $\mathcal{P}_{\zeta}^{\rm CMB}$ (the red dotted line) with the reduced scale $k/k_{\rm p}$, respectively. On the PBH scale where $k/k_{\rm p}\sim 10^{15}$, $\mathcal{P}_\zeta$ needs to be enhanced up to $10^{-3}$--$10^{-2}$ in order to have $f_\pb=1$. Also, ${\cal P}_{\zeta_\phi}$ almost coincides with ${\cal P}_\zeta$ on small scales, meaning that the curvaton does not affect PBH formation significantly. On the scales where $k/k_{\rm p}<10^{12}$, ${\cal P}_\zeta$ is slightly higher than ${\cal P}_{\zeta_\phi}$, indicating that the curvaton can influence large-scale physics (e.g., the primordial non-Gaussianity) and help erase the dip in ${\cal P}_{\zeta_\phi}$. On the CMB pivot scale $k_{\rm p}$, we have $n_{\rm s}=0.9678$, $r_{\rm ts}=0.00250$, and $f_{\rm NL}=0.0555$, all in agreement with the current observations \cite{Planck:2018vyg, Planck:2018jri, BICEP:2021xfz, Planck:2019kim}.}
\label{fig:Pk1}
\end{figure}

The power spectrum $\mathcal{P}_\zeta$ of the total primordial curvature perturbation $\zeta$ as a function of the reduced scale $k/k_{\rm p}$ is illustrated in Fig. \ref{fig:Pk1}. For comparison, we also include $\mathcal{P}_{\zeta_\phi}$ (i.e., the power spectrum only from the inflaton-induced curvature perturbation) and $\mathcal{P}_{\zeta}^{\rm CMB}$ (i.e., the power spectrum extrapolated from the CMB observations on large scales) over the same range. Our basic results are drawn as follows. 

First, on large scales, $\mathcal{P}_\zeta$ aligns closely with $\mathcal{P}_{\zeta}^{\rm CMB}$, but notable deviations appear on the PBH scale as anticipated. This indicates that simple SR inflation alone is insufficient for a large PBH abundance, and a USR phase is necessarily needed. Besides, on the CMB pivot scale $k_{\rm p}=0.05\,{\rm Mpc}^{-1}$, our model yields the scalar spectral index $n_{\rm s} = 0.9678$, the tensor-to-scalar ratio $r_{\rm ts} = 0.00250$, and the local nonlinear parameter $f_{\rm NL} = 0.0555$, all consistent with current observational constraints \cite{Planck:2018vyg, Planck:2018jri, BICEP:2021xfz, Planck:2019kim}. 

Second, on the PBH scale, the near overlap of $\mathcal{P}_\zeta$ and $\mathcal{P}_{\zeta_\phi}$ suggests that the curvaton has little impact relative to the inflaton during the USR phase. However, there is notable difference between these two power spectra on large scales (attention, the vertical axis is in logarithmic scale), indicating a considerable contribution of the curvaton to the total curvature perturbation. As clearly illustrated in Fig. \ref{fig:Pk1}, the curvaton plays a crucial role in enhancing the amplitude of the power spectrum on the scales with $k<k_{\rm PBH}$. A typical feature of this issue is that the dip frequently observed in the power spectrum ${\cal P}_{\zeta_\phi}$ \cite{Zhao:2023zbg, Zhao:2024yzg} (see the blue line around $k/k_{\rm p}\sim 10^{12}$) is absent in our mixed field model (see the black line). The disappearance of this dip is due to the fact that, on this special scale, the contribution from the inflaton almost vanishes, while the contribution from the curvaton keeps almost constant throughout, as shown in Eq. (\ref{Pdtchi}). 

Last and more important, the curvaton also exerts a nontrivial influence on the scalar spectral index $n_{\rm s}$, as indicated in Eq. (\ref{ns}). In other words, since the power spectrum $\mathcal{P}_{\dt}$ induced by the curvaton in Eq. (\ref{Pdtchi}) maintains nearly constant across different scales, it flattens the slope of the total curvature perturbation power spectrum $\mathcal{P}_\zeta$ on large scales. In the absence of the curvaton, from Eqs. (\ref{V1})--(\ref{dtV}), the inflaton potential $V_{1}(\phi)$ predicts a relatively smaller scalar spectral index as $n_{\rm s}=0.9591$, which is even less than its lower limit $n_{\rm s}=0.9607$ at 68\% CL \cite{Planck:2018vyg, Planck:2018jri}. Therefore, the curvaton can be considered as a savior for the inflaton potential $V_{1}(\phi)$, significantly improving its consistency with observations.

\subsubsection{Case 2} \label{sec:case2}

In Sec. \ref{sec:case1}, we have explored the mixed field inflation model with the inflaton potential $V_{1}(\phi)$. Due to the anti-symmetric character of the perturbation $\dt V(\phi)$ on the background potential $V_{\rm b}(\phi)$, $V_{1}(\phi)$ exhibits relatively simple features. Except for $V_0$, other model parameters have little effect on large scales.
%, so the scalar spectral index $n_{\rm s}$ is almost consistent and not excessively unreasonable.
On the contrary, we now investigate an inflaton potential that inherently contains (near-)inflection points, the situation will become much more complicated.

Below, we focus on the (near-)inflection-point inflaton potential $V_2(\phi)$ proposed in Ref. \cite{Garcia-Bellido:2017mdw},
\begin{align}
V_{2}(\phi)=\f{\Lambda}{12}\phi^2\nu^2\f{6-4a{\phi}/{\nu}+3{\phi^2}/{\nu^2}}{(1+b{\phi^2}/{\nu^2})^2}, \label{V2}
\end{align}
where $\Lambda$, $\nu$, $a$, and $b$ are the model parameters \cite{Garcia-Bellido:2017mdw}. The parameter $\Lambda$ sets the amplitude of the power spectrum on the CMB scale, while the others shape the potential. Additionally, to ensure that $\phi$ has only one real solution at the near-inflection point, the parameter $b$ must be slightly less than $1+a^2\{[9/(2a^2)-1]^{2/3}-1\}/3$ \cite{Garcia-Bellido:2017mdw}. For this potential, both the parameters $b$ and $\nu$ can enhance the power spectrum on the PBH scale. However, since $b$ has an upper limit, a sufficiently large $\nu$ is required to compensate for it. Unfortunately, as $\nu$ increases, the scalar spectral index $n_{\rm s}$ decreases. Therefore, to achieve a sufficiently large peak in the power spectrum, it is necessary to include the curvaton to correct the value of $n_{\rm s}$ on the CMB scale.

\begin{figure}[htb]
\centering 
\includegraphics[width=1\linewidth]{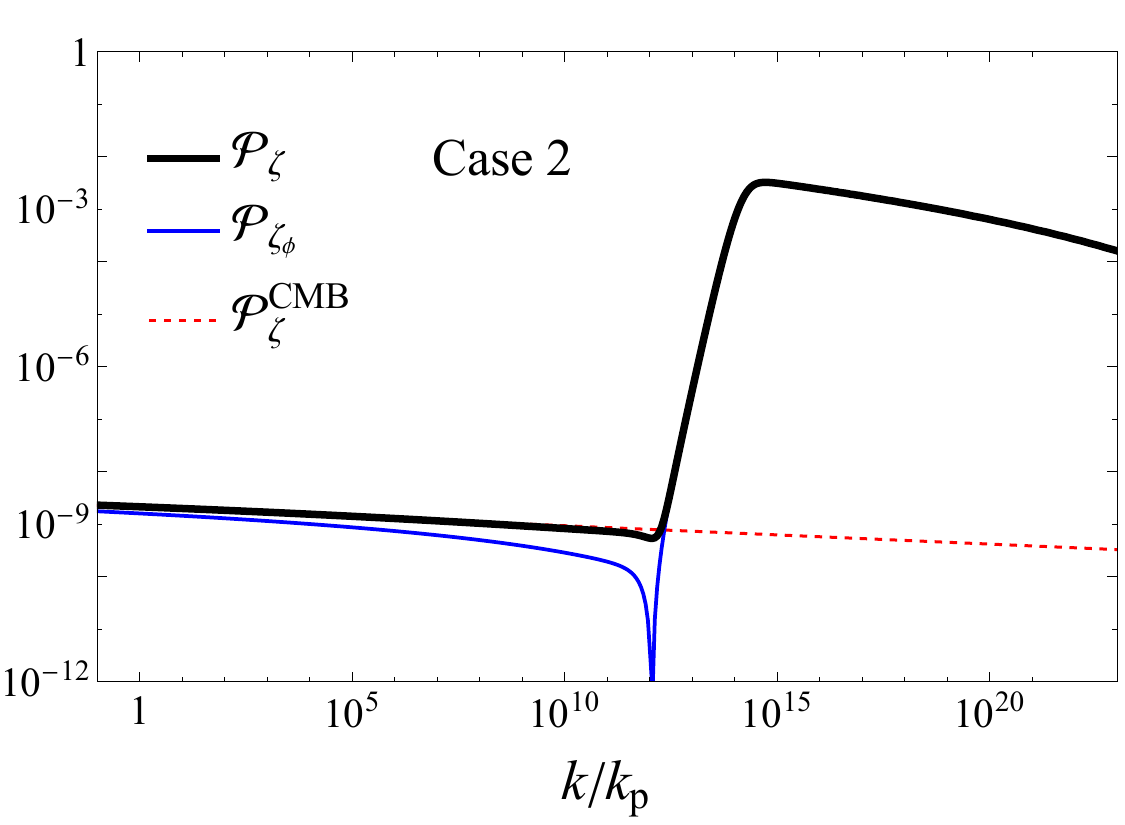}
\caption{
%The power spectra of total curvature perturbation $\mathcal{P}_\zeta$ (black line), of only inflaton-induced curvature perturbation $\mathcal{P}_{\zeta_\phi}$ (blue line), and extrapolated from the CMB observations on large scales $\mathcal{P}_{\zeta}^{\rm CMB}$ (red dotted line) with the reduced scale $k/k_{\rm p}$, respectively. All aspects are similar to Case 1,
Same as Fig. \ref{fig:Pk1}, and we have $n_{\rm s}=0.9670$, $r_{\rm ts}=0.00073$, and $f_{\rm NL}=-0.0520$ on the CMB scale.}
\label{fig:Pk2}
\end{figure}

The model parameters that we employ read $\Lambda=5.43\times10^{-5}$, $\nu/m_{\rm P}=0.041042$, $a=0.2$, $b=1.2955458$, $r=0.8$, and $\bar\chi_*/m_{\rm P}=10^{-2}$, and the initial conditions are set as ${\phi}_{\rm i}/m_{\rm P}=1.01$ and ${\phi}_{{\rm i},N}/m_{\rm P}=-0.0091$. With these parameters, the two resulting power spectra, $\mathcal{P}_\zeta$ and $\mathcal{P}_{\zeta_\phi}$, are displayed in Fig. \ref{fig:Pk2}. The power spectrum $\mathcal{P}_\zeta^{\rm CMB}$ extrapolated from CMB observations on large scales is also included for comparison.
% Similar to Fig. \ref{fig:Pk1}, are plotted with black solid, blue solid, and red dashed lines, respectively.

From Fig. \ref{fig:Pk2}, it is clear to see that the power spectrum $\mathcal{P}_{\zeta_\phi}$ generated merely by $V_2(\phi)$ in Eq. (\ref{V2}) significantly deviates from $\mathcal{P}_\zeta^{\rm CMB}$ on small scales. Moreover, from $V_2(\phi)$ alone, we obtain the scalar spectral index $n_{\rm s}=0.9562$, which is again less than its lower limit $n_{\rm s}=0.9607$ at 68\% CL \cite{Planck:2018vyg, Planck:2018jri}. Fortunately, when the curvaton is included, we are able to obtain $n_{\rm s}=0.9670$, $r_{\rm ts}=0.00073$, and $f_{\rm NL}=-0.0520$ on the CMB scale, which are well consistent with the current observational data \cite{Planck:2018vyg, Planck:2018jri, BICEP:2021xfz, Planck:2019kim}. 

Overall, in the above two mixed field inflation models, the production of PBHs is primarily driven by the inflaton potential $V(\phi)$, while the curvaton field $\chi$ mainly influences large-scale structure and enriches inflationary dynamics, providing the features such as primordial non-Gaussianity. Furthermore, if a single-field inflaton potential fails to meet the observational constraints on the scalar spectral index $n_{\rm s}$, the curvaton can essentially help  correct the models.

\subsection{PBHs from the mixed field inflation model} \label{sec:PBH}

\begin{figure}[htb]
\centering 
\includegraphics[width=1\linewidth]{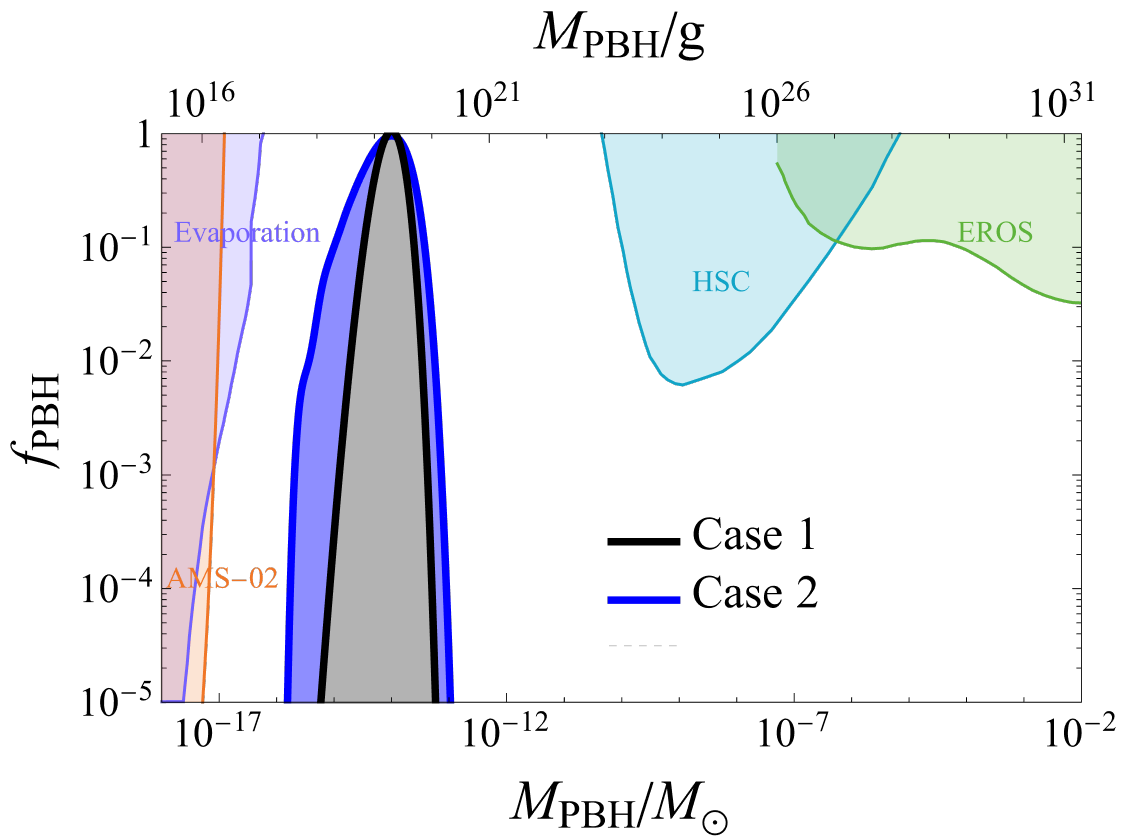}
\caption{The PBH abundance $f_{\rm PBH}$ with the PBH mass $M_{\rm PBH}$ around $10^{-14}\,M_\odot$ for Case 1 (the solid black line with half shade) and Case 2 (the solid blue line with half shade). Here, we demand $f_{\rm PBH}=1$ with our model parameters, so that PBHs can make up all DM. The current observational constraints on $f_{\rm PBH}$ \cite{Carr:2020gox}, including the Hyper Suprime-Cam microlensing (HSC), Exp\'{e}rience pour la Recherche d'Objets Sombres microlensing (EROS), and various evaporation observations (Evaporation) such as EDGESevap, CMBevap, INTEGRAL, 511 keV, Voyager, and EGRB, as well as additional evaporation limits derived from the AMS-02 data in our previous work (AMS-02) \cite{Su:2024hrp}, are also plotted for reference.}
\label{fig:fM2}
\end{figure}

The PBH abundance $f_{\rm PBH}$ with the PBH mass $M_{\rm PBH}$ around $10^{-14}\,M_\odot$ for Case 1 (the solid black line with half shade) and Case 2 (the solid blue line with half shade) are both plotted in Fig. \ref{fig:fM2}. With our model parameters as listed in Sec. \ref{sec:power}, we successfully obtain $f_{\rm PBH}=1$, so as to account for the whole DM. Various observational constraints on $f_{\rm PBH}$ \cite{Carr:2020gox, Su:2024hrp} are also overlaid in different colors for reference. Our results support the production of PBHs in the inflaton--curvaton mixed field inflation model, without contradicting the current observational constraints on $f_{\rm PBH}$. 

In summary, compared to single-field USR inflation, adding a curvaton field introduces richer physical phenomena. On the one hand, the curvaton contributes to the amplitude of the large-scale power spectrum and induces non-negligible primordial non-Gaussianity. On the other hand, it can modify the scalar spectral index $n_{\rm s}$ and save inflaton potentials that not well compatible with the CMB observations on large scales. In addiction, in our mixed field inflation model, we have assumed no direct coupling between the inflaton and curvaton fields for simplicity. If one extends the model with the  coupling, inflationary dynamics will become considerably more complex. This possibility will be left for future exploration. 

\section{Conclusion} \label{sec:con}

PBHs are considered as a natural and compelling candidate for DM as they can form in the very early universe from the collapse of large density fluctuations, avoiding the need to introduce new particle species. This possibility has spurred extensive research on inflation models that can enhance the small-scale curvature perturbation, which is crucial for PBH formation. Nowadays, both single- and multi-field inflationary scenarios have been investigated, with single-field models often incorporating features such as the USR phases to amplify the power spectrum, while multi-field models exploring additional degrees of freedom that can enrich the dynamics of PBH formation.

In this paper, we have developed an inflaton--curvaton mixed field inflation model, a multi-field approach that integrates the USR mechanism commonly used in single-field models. This model not only addresses the need for the enhanced small-scale curvature perturbation, but also allows for a broader exploration of the features such as the primordial non-Gaussianity. In our setup, the inflaton field $\phi$ drives inflation and undergoes a transition from the standard SR phase to a USR phase, amplifying the small-scale curvature perturbation $\zeta_\phi$ necessary for PBH formation. Simultaneously, the curvaton field $\chi$ generates the entropy perturbation ${S}_\chi$ during inflation. After inflation but before BBN, the curvaton decays into radiation, converting the entropy perturbation into curvature perturbation $\zeta_\chi$ and introducing non-Gaussian features into the primordial power spectrum ${\cal P}_\zeta$ of the total curvature perturbation $\zeta$. The relation of $\zeta_\phi$, $\zeta_\chi$, and $\zeta$ can be found in Eq. (\ref{zetabbb1}), and relation of their corresponding power spectra can be found in Eq. (\ref{Pzeta}).

We have further investigated how the model parameters, like $r$ and $\bar\chi_*$, affect the primordial non-Gaussianity. We find that, if these parameters are too small ($r\lesssim 0.2$ and $\bar\chi_*/m_{\rm P}\lesssim 10^{-3}$), the resulting non-Gaussian parameter $f_{\rm NL}$ grows significantly and may even exceed its current observational limits. Taking these constrained model parameters into account and using the $\delta N$ formalism, we can eventually achieve ${\cal P}_\zeta$.
% analyzing both the power spectrum and the non-Gaussian characteristics.

Finally, we have explored two types of inflaton potentials with the USR phases in the framework of our mixed field model, one with a perturbation on the background inflaton potential and the other with (near-)inflection points inherently. By selecting model parameters, our models can successfully produce the PBHs with mass $M_\pb\sim 10^{-14}\,M_\odot$ and abundance $f_{\rm PBH}=1$. Our results demonstrate that on the PBH scale, the influence from the curvaton field is negligible, allowing the inflaton field to dominate PBH formation through the USR mechanism. However, the curvaton field makes a relatively significant contribution to the curvature perturbation on large scales, comparable to the effect of the inflaton. Furthermore, it can even rescue the inflaton potentials that would otherwise produce too low scalar spectral indices $n_{\rm s}$, thus enhancing the viability of these inflaton potentials. With proper parameter tuning, our mixed field model can generate enough PBHs to account for all DM, while remaining consistent with current observational constraints. The synthesis of the USR phase and the curvaton-induced perturbation within the multi-field framework offers a flexible and robust approach for PBH formation, highlighting the importance of such models in current PBH research.

\vspace{.5cm}
\noindent {\bf Acknowledgements} We thank Guansen Wang and Shi-Jie Wang for fruitful discussions. This work is supported by the National Key R\&D Program of China (Grant No. 2022YFF0503304), the National Natural Science Foundation of China (Grant Nos. 12373002, 12220101003, and 11773075), and the Youth Innovation Promotion Association of Chinese Academy of Sciences (Grant No. 2016288). 

\vspace{.5cm}
\noindent {\bf Data Availability Statement} No Data associated in the manuscript.

%\nocite{*}

{\small
\bibliographystyle{spphys}
\balance
\bibliography{ref}
}

\end{document}